\documentclass[universe,review,accept,pdftex,moreauthors]{Definitions/mdpi} 
\firstpage{1} 
\pubvolume{1}
\issuenum{1}
\articlenumber{0}
\pubyear{2024}
\copyrightyear{2024}
\datereceived{6 February 2024} 
\daterevised{7 March 2024} 
\dateaccepted{19 March 2024} 
\datepublished{ } 
\hreflink{https://doi.org/} 

\usepackage{aas_macros,multirow,amsmath}
\graphicspath{{Figs/}}

 \Title{A Needle in a Cosmic Haystack: A Review of FRB \linebreak  Search Techniques}
\TitleCitation{A Needle in a Cosmic Haystack: A Review of FRB Search Techniques}

\Author{Kaustubh 
M. Rajwade $^{1}$ 
 * 
\orcidA{} and Joeri van Leeuwen $^{1}$ \orcidB{}}

\AuthorNames{Kaustubh Rajwade and Joeri van Leeuwen}

\AuthorCitation{Rajwade, K.M.
; \linebreak  van Leeuwen, J.}

\address{%
$^{1}$  \quad ASTRON- 
 The Netherlands Institute for Radio Astronomy, Oude Hoogeveensedijk 4, \linebreak  7991 PD Dwingeloo, The Netherlands 
\\ 
}

\corres{\hangafter=1 \hangindent=1.05em \hspace{-0.82em}Correspondence: kaustubh.rajwade@physics.ox.ac.uk}

\firstnote{Current address: Astrophysics, University of Oxford, Denys Wilkinson Building, Keble Road, Oxford OX1 3RH, UK  } 

\abstract{Ephemeral Fast Radio Bursts (FRBs) must be powered by some of the most energetic processes in the
Universe. That makes them highly interesting in their own right, and as precise probes for estimating cosmological
parameters.
This field thus poses a unique challenge: FRBs must be detected promptly and immediately localised and studied based only on that single
millisecond-duration flash. 
The problem is that the burst occurrence is highly unpredictable
and that their distance strongly suppresses their brightness.
Since the discovery of FRBs in single-dish archival data in 2007, detection software has evolved tremendously.
Pipelines now detect bursts in real time within a matter of seconds, operate on interferometers, buffer
high-time and frequency resolution data, and issue real-time alerts to other observatories for rapid multi-wavelength follow-up.
In this paper, we review the components that comprise a FRB search software pipeline, we discuss the proven techniques that were adopted from pulsar searches, we highlight newer, more efficient techniques for detecting FRBs, and we conclude by discussing the proposed novel future methodologies
that may power the search for FRBs  in the era of big data astronomy.
}


\begin{document}

\section{Introduction}

The universe has fascinated humans since they first gazed into the sky and started to wonder. The advancements in
science and technology have meant that we can now even see the cosmos through previously invisible light such as radio
waves. And, while the everyday sky appears mostly static, it turns out to be highly variable and transient at
wavelengths our eyes are oblivious to. One such transient phenomenon is Fast Radio Bursts (FRBs). These are bright radio flashes that travel from billions of light years away. The measured fluxes and duration suggest a highly energetic and ephemeral physical process. These attributes make FRBs a key tool for answering outstanding questions in cosmology and probing the most energetic processes in the universe.

FRBs were serendipitously discovered in 2007 when astronomers were mining vast amounts of radio data at the
highest time resolution for pulsars, radio-emitting neutron stars~\citep{lorimer_2007_sci}. Since the discovery of
pulsars in 1967~\citep{hewish1968}, astronomers have developed various techniques to look for celestial radio
pulses.  
Telescopes find radio pulsars either by a periodicity search or by detecting individual bright radio
pulses~\citep{lorimer_2012_hpa}. By the early 2000s, more than 2000 radio pulsars had been discovered, and  an
increasing effort was ongoing to find more in our Galaxy and in the Magellanic clouds~\citep{lorimer2006,
  manchester2006}. One the main things to correct for while searching for bright narrow radio pulses is the dispersion
of the radio waves by the interstellar medium. This causes radio signals at lower frequencies to reach Earth later than their counterparts at higher frequencies, manifesting itself as a sweep of the pulse across the observing band. The amount of
dispersion depends on the total electron column density along the line of sight; this measure thus acts as a proxy
for the distance between the source and the observer. Astronomers search for pulses over a range of dispersion values,
up to the largest distance they want to cover. Until 2007, no searches were carried out for bright pulses beyond the
local Galactic neighbourhood, while searches in nearby galaxies were sporadic~\citep{manchester2006}. All that was about to change
when an extremely bright burst was discovered in 2007, with a dispersion much larger than what was expected from the
Milky Way, in archival data from the Murriyang (Parkes) telescope. This seminal event in time-domain astronomy opened up a new parameter space of cosmological radio bursts with tremendous scientific potential~\citep{macquart_2020_natur}.

The discovery made the community
realise that there are regions on the transient phase space that remain unexplored because (1) they are selected against
in our surveys and (2) we did not expect any sources in those regions. Recent discoveries of FRBs and other radio
transients have proven point 1 to be true, but point 2 much less so---highlighting the importance of revising our data
processing techniques and conducting sensitive searches for fast radio transients with current and future
telescopes. In this review, we provide a summary of the past and present techniques used by radio astronomers
to find FRBs in time-domain data. We go through the search methodologies and discuss the advantages and the
underlying biases. We also discuss some newer techniques that are currently being deployed to expand our horizons
within the radio transient phase space and we provide an outlook towards the future of FRB searches with next-generation radio telescopes.

\section{Key Steps in a FRB Search Pipeline}
\begin{figure}[H]
\begin{adjustwidth}{-\extralength}{0cm}
\centering
\includegraphics[width=18.5cm]{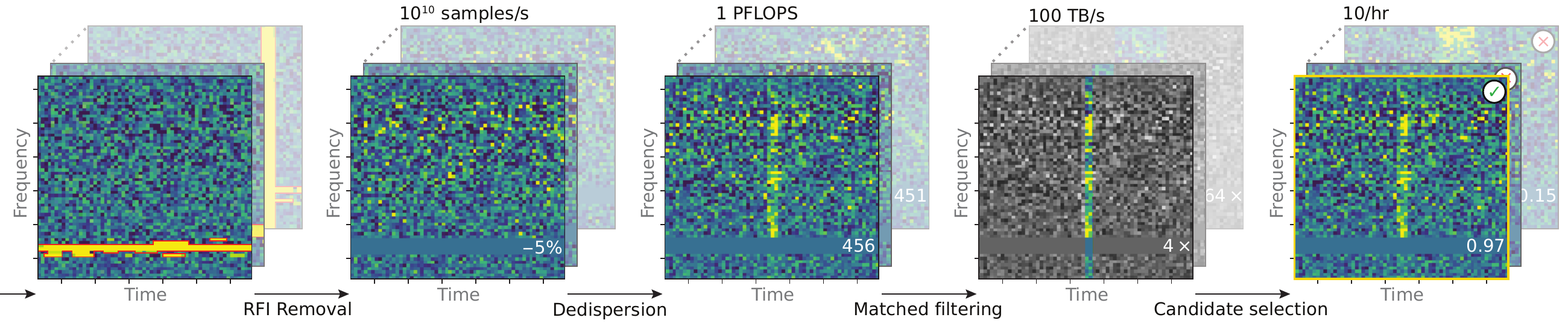}
\end{adjustwidth}
\caption{The various steps required in a FRB search and their impact on the time-frequency plane. 
Fiducial rates and parameter values are taken from \citep{2023A&A...672A.117V}. 
Listed at the top for each step are the relevant data and compute and candidate rates. 
From left to right, data arrive from the telescope. 
These are cleaned from RFI. 
Order 10$^4$ dispersion trials are next formed and each is searched over $\sim$10~matched filters. 
The large number of candidates are next clustered and graded until $\sim$10$^3$ remain, of which one is a new FRB.
The FRB in this example was found after 5\% data loss to RFI, at a DM of 456~$\rm pc\,cm^{-3}$ 
 and a downsampling factor of 4. 
\label{fig:steps}
}
\end{figure}

The survey technologies employed at present reduce and analyse roughly ten Exabyte of raw data to find a single FRB (e.g.,~\citep{2023A&A...672A.117V}). 
This analysis can be thought of as a number of distinct functional steps, even if these are sometimes optimised into one
another at deployment. 
This data analysis pipeline has to process large volumes of data, but has strong constraints on latency, too. To fully utilise the scientific potential of FRBs, it is important to localise the bursts by imaging the radio
sky at the time of the FRB and sending alerts to other observatories. This can help constrain any multi-wavelength
emission that may be concurrent to the event. This is only possible if both the FRB is detected and the corresponding
radio data are saved in real-time. This means that FRB detection software must be extremely efficient and exceedingly
 fast. This has led to vast developments in making every component within the FRB search methodology highly optimised. We illustrate this sequence in Figure~\ref{fig:steps}
 and describe its parts in detail below. We provide a (not exhaustive) list of publicly available software that implements all the steps described below in Appendix~\ref{app1}.
This sequence produces and searches an order of $10^{12}$ realisations of the data for every FRB that is discovered.
It is instructive to compare this stupendous number against the odds of finding the proverbial 
needle in the haystack. To match the effort required in a FRB search,
the grassy pile would need to rival the Great Pyramid of Giza.

\subsection{Rejection of Radio Frequency Interference}

Whenever astronomers have looked at the cosmos through radio wavelengths, they have been greeted by a large diversity of radio
signals created by humanity. These signals, while undoubtedly important for the progress of civilisation, are a nuisance
in detecting any faint radio signal from the Universe. This so-called ``radio frequency interference'' (RFI)
can be broadly classified into two distinct categories: broad-band and narrow-band RFI. Broad-band RFI refers to the spurious signals that affect the entire band of observations. Narrow-band RFI is limited in its spectral coverage and can be caused by communications at specific frequencies or any strong periodic signal that is terrestrial in nature. Any signal that is not astrophysical is treated as RFI in radio astronomy. Hence, various tools and techniques have been devised over the last 50 years to reduce the deleterious effects of RFI on fainter signatures from FRBs.

A key feature that differentiates an astrophysical radio signal from a terrestrial one is the dispersion
measure (DM).
As radio waves travel through the intergalactic and interstellar plasma, they encounter free electrons that act
like a refractive material, causing radio waves at lower frequencies to be delayed with respect to the higher
frequencies. This manifests itself as a delay at the receiving antenna, observed at a band between $\nu_1$ and $\nu_2$~as
 \begin{equation}
    \Delta t = 4.15\,\left(\nu_1^{-2} - \nu_2^{-2}\right)\, \rm DM\, ms,
    \end{equation}
    where,
    \begin{equation}
    {\rm DM} = \int_0^l n_e~dl.
    \end{equation}
 
 Here, $n_e$ is the electron density and $l$ is the distance to the source. The DM is a measurable quantity that
 is proportional to the distance to the source. Astronomers can use this attribute to identify RFI, as any terrestrial
 or satellite RFI will have a DM of 0 $\rm pc\,cm^{-3}$. Below, we list a few of the popular RFI excision techniques used in time-domain radio data,
 ordered by the level of complexity.

\subsubsection{Static Channel Mask}
Any radio telescope has a number of persistent, local RFI sources. Some of these sources are always emitting at the same
radio frequency with little time variability. In these situations, it is  sensible to pre-emptively mask out the
relevant frequencies from the telescope data as they will be completely unusable. Hence, astronomers can generate a
static frequency channel mask for these known sources of RFI that are always present in the data. One of the main
drawbacks of a static mask is that it cannot account for the time variability of RFI, or for the corruption of a new
region of the frequency band. Then, the mask needs to be updated. Hence, more sentient RFI flagging algorithms are used
for telescopes around the world nowadays. 

\subsubsection{Zero-DM Filter}
The zero-DM filter first de-disperses the time series at a DM of 0 $\rm pc\,cm^{-3}$ and then subtracts these values 
from every frequency channel. Any broadband RFI that exists in the data---which is strongest
at a DM of 0 $\rm pc\,cm^{-3}$, while any astrophysical signal is expected to have a positive, non-zero DM---is mostly
removed.
While this recipe is commonly used in standard searches for FRBs~\citep{eatough2009}, there are certain
drawbacks.
This is especially the case for sources with a high signal-to-noise ratio (S/N),
that can be detected in every frequency channel.
As we are effectively subtracting the  mean of the channels from the time-observing frequency intensity data from the telescope (henceforth called a dynamic spectrum),
and these high-S/N cases, one also removes
some amount of the true signal. This results in a reduction in the significance of the signal. Secondly, for sources
that have very small DMs, the dispersed signal resembles broadband zero-DM RFI and such a subtraction can remove a
significant amount of astrophysical signal. But, overall, since FRBs are expected to have very large DMs, and the S/N per
channel is usually low, this method is a simple and effective tool for  search engines.

\subsubsection{Real-time Thresholding}
Typical time-domain surveys contain a sequence of time-frequency data recorded with integration times of a few tens to
hundreds of microseconds. At these timescales, the RFI environment at a radio telescope changes rapidly, which makes a
static channel mask ineffective for real-time searches for FRBs.
FRB surveys require such real-time processing as the high data rates preclude storing the full-resolution observations for a meaningful length of time.
This means RFI mitigation needs to occur on-the-fly as well.
This is a challenge because
(1) the tension between hardware memory limitations and the varying but ever-incoming data stream mean the
algorithm has only limited access to the data statistics that change over time
and (2) the compute bounds and real-time requirement demand a simple algorithm. 
In practice, for 
 a multi-beam survey in which 10$^2$\,Gbps of incoming data need to be cleaned of RFI in
  real-time, as in Figure~\ref{fig:steps}, 
  these requirements translate to computer demands that
  require $\sim$1--5\% of the overall execution time \citep{sclocco_real-time_2020,2023A&A...672A.117V},
  operating on the local statistics of buffered data chunks spanning $\sim$1\,s.
The FRB surveys at, e.g.,~CHIME~\citep{2023ApJS..265...62R} and Apertif~\citep{sclocco_real-time_2020}
each employed algorithms that identify and remove
samples stronger than a defined noise threshold
and that adapt dynamically to the noise characteristics of the
incoming data. These can act in both time and frequency domain
and can be applied iteratively.
The {\tt RFIm}  real-time mitigation library~\citep{sclocco_real-time_2020}
that implements this algorithm is open source.

\subsubsection{IQRM}
A method that more explicitly removes slowly varying trends in the data than the thresholding described above, and
detects RFI outliers, is the Inter-Quartile Range Mitigation (IQRM) algorithm. IQRM identifies all the frequency channels affected by narrow-band RFI. The first step is to characterise the entire chunk of data by a single statistical metric per frequency channel and then to determine the channels whose statistical
metric does not conform to clean data. IQRM is characterised by two aspects. First, any spectral statistic to decide
which channels are contaminated by RFI is acceptable, as long as higher values of the metric indicate a larger
contamination of the data by RFI. Second, the threshold beyond which data are rejected is estimated from data which are assumed to vary slowly with the observing frequency.

The algorithm works as follows: for each sample in a frequency channel $i$, we compute the chosen statistical metric $s^{0-N}_{i}$. The total number of samples considered are dictated by a radius parameter and a threshold parameter (in units of $\sigma$) that the metric needs to exceed to be considered an outlier. Using these parameters, the algorithm computes the difference, $\Delta~s^{j}_{i}$, over the radius $r$ such that 0 $\leq~j\leq~r$. Finally, all the $\Delta~s^{j}_{i}$ values are filtered based on the threshold defined by the user. Any values above this threshold will be flagged as RFI by the algorithm.

\begin{figure}[H]
\begin{adjustwidth}{-\extralength}{0cm}
\centering
\includegraphics[width=14.5cm]{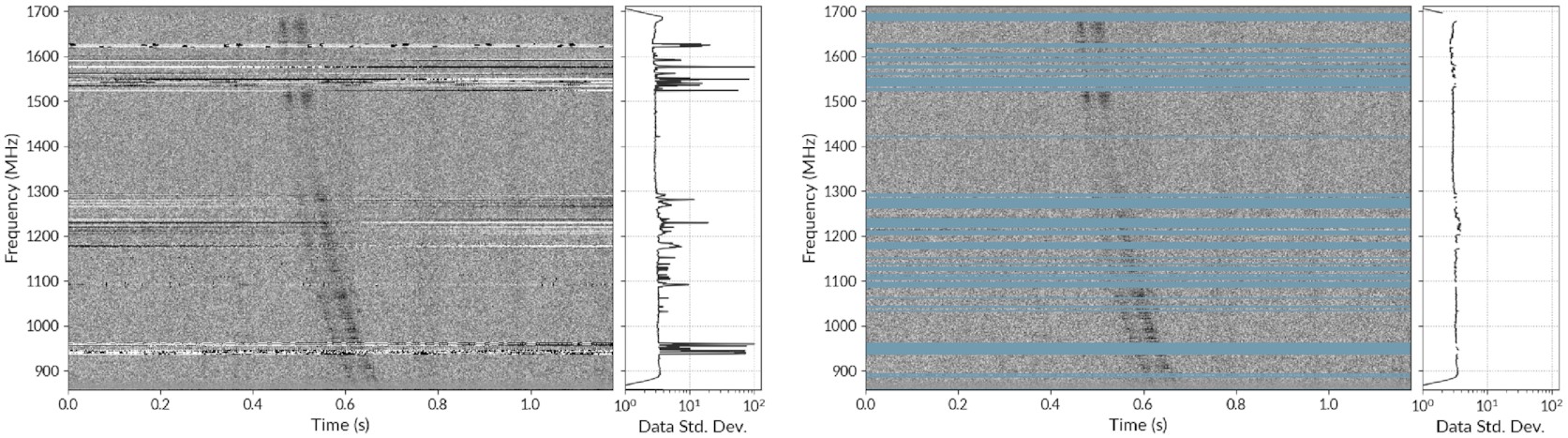}
\end{adjustwidth}
 \caption{A dynamic spectrum 
 at 1.4~GHz taken with the MeerKAT telescope 
 with a single pulse from a PSR J1226-3223. The
  left panel shows the spectrum before any RFI excision and the right panel shows the spectrum after running IQRM on the
  data. The figure has been adopted from Figure~3\mbox{ in~\cite{morello2022}.}\label{fig:iqrm}}
\end{figure}

 Figure~\ref{fig:iqrm} shows an example of data before and after flagging by IQRM. It clearly demonstrates the ability
 of IQRM to pick out time variable narrow-band RFI. IQRM has been used for various time-domain surveys~\citep{morello2022}. Since the algorithm is dependent on a statistical metric that is expected to be robust
 to RFI, it is bound to be ineffective in cases where RFI affects a significant fraction of the data that is used to
 compute the statistical metric. Hence, IQRM is not an effective tool against long-term, broad-band RFI.
 
\subsubsection{Z-dot Filter}
The Z-dot filter is an updated zero-DM filter that computes the zero-DM time series similar to the zero-DM filter, but subtracts only the corresponding contribution from each  channel. This avoids over-subtraction in channels where there is a real signal and results in improved sensitivity of the search.

Similar to the zero-DM filter, the zero-DM time series is estimated by de-dispersing the original data at zero-DM
value. Let the corresponding time series be denoted by $\rm s_{dm = 0}$. One then estimates 
the contribution per channel by determining the baseline $\beta_{i}$ plus the scale factor $\alpha_{i}$ 
such that the residuals for fitting the zero-DM 
contribution are minimised for each
channel $i$.  Hence, a reduced-$\chi^{2}$
metric is  defined such that,
\begin{equation}
\chi^{2} = \left({\rm t_{i}} - \alpha_{i}{\rm t_{dm=0}}  -\beta_{i}\right)^{2}.
\end{equation}
where $\rm t_{i}$ is the time series of the i\emph{th} channel and 
\begin{equation}
\alpha_{i} = \frac{{\rm t_{dm=0}\cdot t_{i}} - \frac{1}{N} \sum_0^N {\rm t_{i}}\sum_0^N{\rm t_{dm=0}}}{{\rm t_{dm=0}\cdot t_{dm-0}} - \frac{1}{N} \sum_0^N {\rm t_{dm-0}}\sum_0^N{\rm t_{dm=0}}},
\end{equation}
where the summation is the sum over $N$ time samples in the time series. The final \linebreak  filtered time series,
\begin{equation}
{\rm t^{'}}= {\rm t_{i}} - \alpha_{i}{\rm t_{dm=0}}.
\end{equation}

\textls[-15]{$\beta_{i}$ is just the DC offset in this case and does not need to be subtracted during the filtration process. A big advantage of this method is that the samples are subtracted with different weights, which allows the user to still preserve the full astrophysical signal while effectively removing most of the zero-DM RFI. The full recipe for this technique is provided in~\cite{men2019}.}

\subsubsection{Spectral Kurtosis}
Tests of statistical moments for a certain distribution can serve as a powerful metric for identifying RFI.
Sources in nature typically produce electric fields that are Gaussian, and any extreme non-Gaussianity that rises from
observed RFI sources leads to deviations in the statistical properties of radio measurements. Spectral Kurtosis (SK) is a statistical parameter that can be used to differentiate between
Gaussian noise and non-Gaussian features in the dataset. Kurtosis is the fourth moment of a normal variable. It characterises the sharpness of the  Gaussian distribution~\citep{nita2016}. Typically, stationary broadband astrophysical signals have Gaussian-like kurtosis. The presence of RFI will significantly affect the kurtosis of the data, thus providing astronomers with a way to filter RFI without affecting the signal of interest. From the definition of kurtosis, one can derive the spectral kurtosis estimator, 
\begin{equation}
\Hat{SK} = \frac{XYc + 1}{X -1}\left(\frac{X\cdot A_{2}}{A_{1}^{2}} -1 \right),
\end{equation}
where $X$ is the number of spectral values that are accumulated, $Y$ is the total number of spectra that are used to estimate this value, $c$ is a shape parameter of the response of the detector that is empirically derived, and
\vspace{-6pt}\begin{equation}
  \begin{split}
    A_{1} = \sum_{j=1}^X\left(\sum_{i=1}^Y I_{i}\right)_{j} \\
    A_{2} = \sum_{j=1}^X\left(\sum_{i=1}^Y I_{i}\right)^{2}_{j}
\end{split}
  \end{equation}
  are the first and second statistical moments of the data.
For Gaussian noise, the distribution of $\Hat{SK}$ is predictable and any deviations from it can be identified up to a
certain threshold and confidence and are removed from the data. Several studies have assessed the effectiveness of spectral kurtosis in searches for radio transients~\citep{smith2022, purver2022}.

\subsubsection{Secondary, Off-line RFI Mitigation}
In buffered or off-line data, more sophisticated methods can be used.
With these, FRBs that were detected from the real-time pipeline can be cleaned from RFI to achieve a higher signal-to-noise ratio (S/N).
But, these also reduce the S/N down to which new FRBs can be found and studied.
Off-line, RFI filters can take multiple passes to, e.g., first assess the noise levels in the entire data set,
and next define individual bright outliers as RFI (e.g.,~\citep{2011ascl.soft07017R,2012AR&T....9..237V}), or even
remove low-level periodic RFI throughout \citep{Maan2020}.

\subsection{Dedispersion}
As described in the previous section, radio waves travel from the FRB event to the observer, and traverse cold plasma
(in the host galaxy, the inter-galactic medium and the Milky  Way). A narrow pulse from a FRB is smeared out over the
observing band by an amount depending on the magnitude of the DM, resulting in a sub-optimal detection. Hence, before
performing any search for FRBs in radio data, the data have to be corrected for this dispersive delay.
This process is known as dedispersion.
There is an approximate ``incoherent'' method and an exact ``coherent'' method (Section~\ref{sec:semicodedi}).
Since we have no a priori information about the distance of the FRB, the data have to be dedispersed over a range of trial DMs. Below, we describe some of the key dedispersion algorithms.

\subsubsection{Brute Force}
\label{sec:bf}
The direct dedispersion algorithm sums the intensities of a dynamic spectrum over the entire bandwidth, along the
quadratic sweep corresponding to the trial dispersion measure. The algorithm calculates a dedispersed time series
$t_{{\rm D,t,}\nu}$ from an input dataset $\mathcal{I}$ where the subscripts indicate a given DM $D$,
 time $t$, and frequency
$\nu$. The algorithm is run over a number of frequency channels $N_{\nu}$, number of time samples $N_{\rm t}$, and  a
number of DM trials $N_{\rm D}$. Typically, on modern heterogeneous architectures, the brute force dedispersion
algorithm is parallelised over $N_{\rm t}$  and $N_{\rm D}$, with the summation over $N_{\nu}$ happening serially. One
drawback of the brute force method is that the complexity of this algorithm is very high, $\mathcal{O}(N_{DM}, N_{t},
N_{\nu})$. The other issue with this algorithm is that it requires a large memory bandwidth in the compute hardware and
has a very low arithmetic density once the data are loaded to memory. 
While these challenges appear daunting,
three methods are combined in practice to remedy both issues
(see, e.g.,~\citep{2023A&A...672A.117V}),
allowing us to dedisperse data in real-time.
For the first two, an error budget is estimated once, at the start of a survey,
based on the intra-channel dispersion that is unavoidable for incoherent dedispersion at higher DMs,
and this is encoded in a look-up table.
Then, first, for higher DMs,
the input data are downsampled such that they remain below this error, reducing $N_{\rm t}$ without loss.
Second, dedispersion takes place using a two-step sub-banding algorithm---i.e., first dividing
the entire bandwidth over multiple sub-bands, next dedispersing within these,
and finally combining all sub-bands---effectively reducing  $N_{\nu}$  (cf.~\citep{Barsdell2012}). 
Finally, memory access is tuned such that the required data can be very rapidly retrieved from the cache (``data re-use'',~\citep{sclocco2016}).

\subsubsection{Tree Dedispersion}

\begin{figure}[H]
  \includegraphics[width=7.6cm]{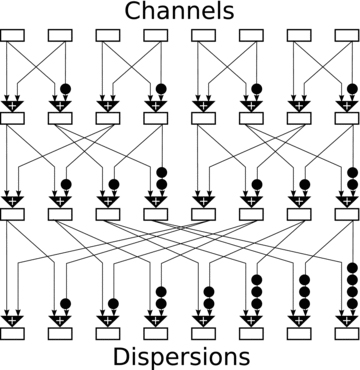}
  \vspace{-1pt}
  \caption{\textls[-15]{The tree dedispersion algorithm that uses the divide and conquer technique to remove redundant operations during dedispersion over many DM trials. The Figure has been adopted \mbox{from~\cite{Barsdell2012}.}}\label{fig:dedisp} }
\end{figure}

Brute force dedispersion can get computationally intensive very quickly for modern day FRB surveys
given their large data volumes and high number of trial DMs. The tree dedispersion algorithm~\citep{taylor1974} uses the
divide and conquer algorithm (similar to the Fast Fourier Transform,~\citep{fft}) to reduce the complexity of
dedispersion. The complexity in this algorithm reduces from $\mathcal{O}(N_{\rm t} N_{\rm D} N_{\nu})$ to
$\mathcal{O}(N_{\rm t} N_{fdm\nu} log(N_{\nu}))$. This significant speed-up is obtained by first regularising the
problem (represent the mathematical problem into a super-position of linear terms) and then eliminating all redundant additions. In spite of the large speed-ups in the calculation, there are
overheads involved with the many add operations which still make this algorithm suited for many-core and GPU (Graphics processing unit) architectures. In a mathematical sense, the algorithm computes the following: 
\begin{equation}
\mathcal{I} (t)= \sum_{\nu}^{N_{\nu}} B(\nu, t + \delta t(d^{'}, \nu)),
\label{eq:tree}
\end{equation}
where $\mathcal{I} (t)$ is the intensity of a given time sample, $B$ is the intensity for a given time sample and observing frequency $\nu$, and
\vspace{-6pt}
\begin{equation}
\delta t(d^{'}, \nu) = d^{'}\left(\frac{\nu}{N_{\nu}-1}\right),
\end{equation}
\textls[-15]{for a bandwidth with $N_{\nu}$ and a DM trial $d^{'}$ in the range 0 $\leq$ $d^{'}$ < $N_{\nu}$. 
This regularisation produces a delay function $\delta t(d^{'}, \nu)$ that is a linear function of frequency. Hence, the DM can also be mapped to a linear function of frequency. 
Due to this regularisation, Equation~(\ref{eq:tree}) can be computed for different DMs in $log_{2}{N_{\nu}}$ steps using the
divide and conquer algorithm (as illustrated in Figure~\ref{fig:dedisp}). While the algorithm reduces the number of
computations, there are several constraints introduced by the method.
The two key ones are that the DMs for which the data are dedispersed are not
the efficiently distributed DM trial values that produce the most sensitive search,
and that the algorithm sums over linear tracks. 
To overcome these, tree dedispersion can first run within subbands (cf.~Section~\ref{sec:bf})
and next we can combine their outcomes with the same algorithm.
That way, tree dedispersion approximates the quadratic DM smear over the entire band while still preserving the efficient computation with regularisation. }

\vspace{-6pt}

\begin{figure}[H]
\includegraphics[width=0.9\textwidth]{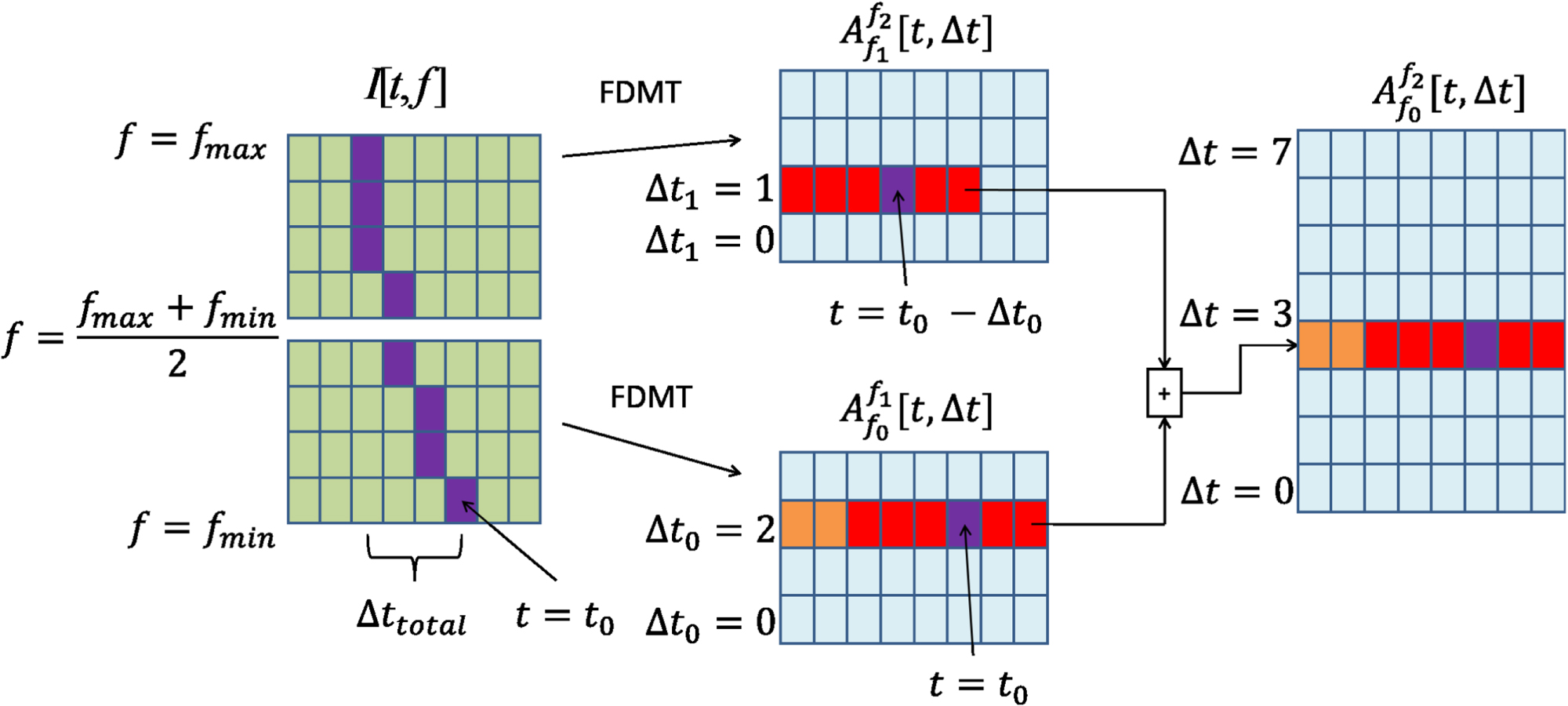}
\caption{An overview of the FDMT technique for 
  dedispersion during FRB searches. FDMT maps the time-frequency plane ($I$[$t$,$f$], \textbf{left}) to the plane (\textbf{right})
  of time versus time-delay
  (a proxy for dispersion measure), $A$[$t$,${\Delta}t$]. The purple DM track (\textbf{left}) is transformed into the single purple pixel~(\textbf{right}). Figure taken from~\cite{zackay2017}.\label{fig:fdmt}}
\end{figure}

An algorithm that also recursively breaks down the problem, like tree dedispersion, is the  
fast dispersion measure transform (FDMT; \citep{zackay2017}).
FDMT (Figure~\ref{fig:fdmt}\endnote{Zackay B., et al., An Accurate 
 and Efficient Algorithm for Detection of Radio Bursts with an Unknown Dispersion Measure, for Single-dish Telescopes and Interferometers, 2017, Astrophysical Journal, Volume 835, Issue 1, DOI: \url{10.3847/1538-4357/835/1/11}. \copyright~AAS reproduced with permission.}) combines a low theoretical complexity as seen in  tree dedispersion with
the ability to sum over the exact quadratic dispersion curve as featured in brute force dedispersion.
The latter is important especially when using high fractional bandwidths, high DMs, and/or low observing frequencies.
While brute force dedispersion implementations successfully employ data reuse to allow real-time searching
\citep{sclocco2016}, and the problem is thus solved in practice,
FDMT is intrinsically more efficient.
\vspace{-6pt}

\subsubsection{Semi-coherent Dedispersion}
\label{sec:semicodedi}
Interstellar dispersion can also be thought of as an impulse response of the ISM as the electro-magnetic wave moves through it. If we are able to measure the electric field at the telescope, it can be written as 
     \begin{equation}
     \mathcal{S}(t) = \mathcal{S_{0}}(t) \ast \mathcal{H}(t),
     \end{equation}
     where $\ast$ denotes a convolution operation and $\mathcal{H}(t)$ is the impulse response function of the ISM such that
     \begin{equation}
         \mathcal{F}(\mathcal{H}(t)) = e^{-i\frac{2\pi~D}{(\nu_{0} + \nu_{d})\nu_{0}^{2}}~DM~\nu_{d}^{2}} ,
     \end{equation}
 where $D$ is a constant, $\nu_{0}$ is the lower edge of the band, $\nu_{d}$ is the deviation from the lower edge, and
 $\mathcal{F}$ is the Fourier transform.
 In the frequency domain, the ISM response function adds an additional phase to
 the signal which can be removed if the DM is known.
 This removal method,  coherent dedispersion, fully corrects for the dispersion with no intra-channel smear. 
 It is implemented by performing 
 a complex multiplication by the inverse of the impulse function in the frequency domain.
 Typically, searches for FRBs are done by performing incoherent dedispersion over a number of DM trials where the
 intra-channel dispersion smear reduces the sensitivity of the search. In ``semi-coherent dedispersion''
 one can use
 coherent dedispersion over a coarse grid of DM trials, and next perform a finer DM search for FRBs incoherently,
 improving the overall sensitivity \citep{2017A&C....18...40B}.

\subsubsection{Fourier Domain Dedispersion}
Typically, brute force dedispersion is performed on time-domain data, where the dynamic spectrum is shifted in time per frequency channel depending on the DM, to generate a new dynamic spectrum. This algorithm requires a large memory bandwidth for only a minimal number of computations. In the Fourier Domain, a time delay corresponds to a phase rotation. These phase rotations can be represented as dense matrices that can be efficiently solved on the GPU/CPU.

In Fourier Domain dedispersion, one can convert the dynamic spectrum $I(t, \nu)$, where $t$ is time and $\nu$ is the observing frequency, to a Fourier spectrum where $k$ defines the Fourier frequencies. This amounts to $N_{\nu}$ ($N_{\nu}$ is the number of frequency channels across the bandwidth) 1-D Fourier transforms $I(t,\nu) \leftrightarrow \mathcal{I}(t, k)$ that should not be confused with a 2-D Fourier transform of the whole dynamic spectrum. Under this framework, the DM delays in the $k$ space can be represented as phase rotations such that
\begin{equation}
\mathcal{A}(k, \nu, {\rm DM}) = e^{-2\pi i k\delta t(\nu, DM)}.
\end{equation}

Hence, the final dedispersion product is 
\begin{equation}
I(t,DM) = \sum_{0}^{N_{\nu}} \mathcal{F}^{-1} (\mathcal{A}(k)~\mathcal{I}(t, k)),
\end{equation}
where $\mathcal{F}$ is the inverse Fourier transform operator. Since the Fourier transform is linear, the inverse transform and the summation operations can be swapped, such that one has to perform the inverse transform only once, thus significantly improving the efficiency of the algorithm. Details of this algorithm are provided in~\cite{bassa2022}.

\subsection{Matched Filtering}
In time-domain signal processing, matched filtering is an effective way of unearthing a non-Gaussian signal from noisy
data if the shape of the expected signal is known a priori. In reality, any time series
\begin{equation}
    X[t] = a[t] + n[t],
\end{equation}
where $a[t]$ is the signal and $n[t]$ is the noise at any given time $t$. Matched filtering involves applying a filter
to the time series such that maximum S/N will occur when the impulse response of the filter is equal to
the shape of the signal in the time series. Mathematically, this means that the filter is convolved with the time series
to detect the signal and determine its physical properties. In time-domain astronomy, this filtering technique is used
widely to detect FRBs and single pulses in noisy data, in real-time. In the simplest case, a matched filter 
is a set of box cars with varying widths which approximates the shape of a single pulse. 
For any time series produced after dedispersing the data at a given DM, these boxcars of varying widths are convolved with the time series and the result is searched for any points above the signal-to-noise threshold. In reality, any real pulse can be best approximated by a Gaussian pulse. Hence, for a time series $X[t]$, the maximum S/N is
\begin{equation}
        S/N_{\rm max} =  {\rm max}\left[X[t] \ast B[t,w]\right],
\end{equation}
where $ B[t]$ is the boxcar function and $\ast$ denotes a convolution operation. $S/N_{\rm max}$ will maximise when the
width $w$ is closest to the true FRB width. Since this width is unknown a priori, the time series has to be convolved
with a large number of boxcar trials, thus increasing the complexity of the search. To optimise this process,  boxcar
widths are geometrically spaced in powers of 2 to convolve the data. The data themselves are also downsampled for larger widths to reduce the data volume/number of time samples and hence, the time of processing. 
Typical widths of FRB range from a millisecond to a few tens of ms~\citep{petroff_2019_aarv}. Hence, real-time
FRB surveys generally only search for transients spanning between an order 0.1 and 100 ms---although recent discoveries have shown the importance of going to larger widths in the search~\citep{caleb2022,hw2022,hw2023}.
    
\subsection{Candidate Classification}
All real-life FRB detection pipelines produce many spurious candidates due to various terrestrial sources. These
so-called 'false positives' make an actual detection extremely challenging in high RFI environments.
Ever-proliferating cellular network towers, radar stations and satellite constellations emitting intense radiation
across the radio-band can generate millions of false FRB candidates in any telescope even after proper RFI excision. Hence, one has to use clever techniques to identify a real astrophysical source from \mbox{the noise}.

\subsubsection{Sifting and Clustering}
\begin{figure}[H]
\centering
\includegraphics[width=\textwidth]{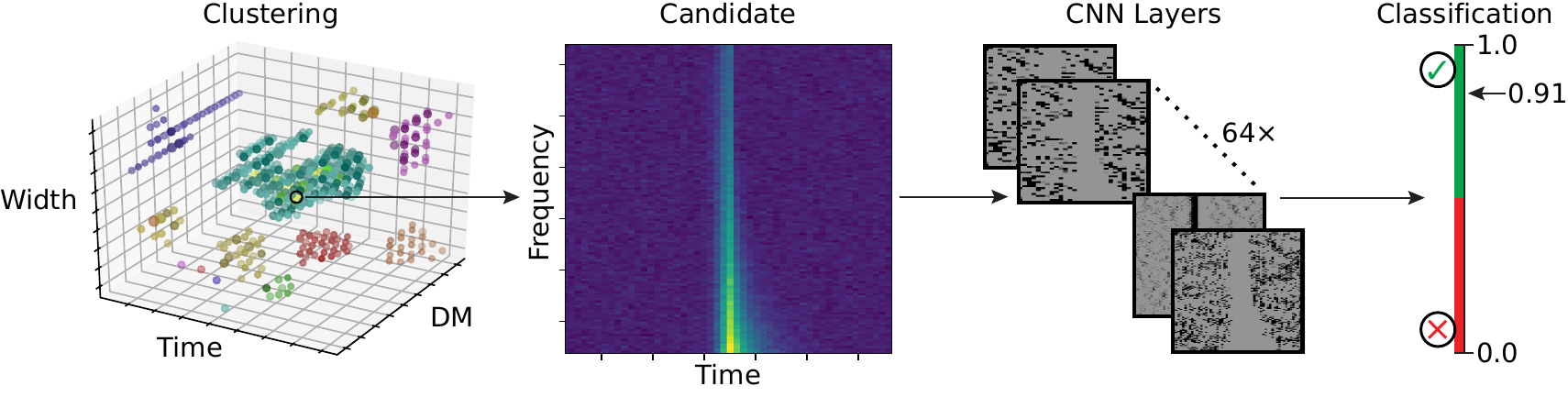}
\caption{Candidate classification generally involves, from left to right, candidate clustering, extraction of best candidates per cluster as a dynamic spectrum, and analysing each of these using a CNN that produces a classification metric.
Each cluster of candidates is marked with a different colour. 
Figure partially adapted from \citep{2018AJ....156..256C}, where  
two convolutional layers (32/64 kernels each), two pooling layers, and two fully connected layers comprise the CNN. 
\label{fig:ml}}
\end{figure}

When time-domain data are searched for FRBs, the search is done over a number of DM and width trials, over all the time
samples. Typically, candidates generated by RFI will all have DMs close to zero. Similarly, real astrophysical signals
of interest do not last for more than 100~ms; hence, any pulse wider than that would most likely have a terrestrial
origin. One can thus introduce cut offs on the DM and widths, such that candidates below a certain DM and above a
certain width are automatically ignored by the real-time pipeline. Such sifting has proven to be an effective way to
reduce the number of spurious hits. Similarly, any strong signal will be detected above the S/N threshold of the
pipeline over a number of time samples, DMs and widths. One can use this property to effectively cluster events and
relate them to a single source/candidate. Popular clustering algorithms include the friend-of-friends algorithm~\citep{press1982},
whereby candidates within a normalised distance threshold in time, DM, and width parameter space from a given candidate
are clustered together (left panel of Figure~\ref{fig:ml}). 

\subsubsection{Machine Learning Techniques}
Even after employing effective candidate sifting and clustering strategies, strong RFI environments can still result in
a large false positive rate, thereby increasing the latency of a real-time FRB detection pipeline. This is crucial as
often, in the event of a detection of a FRB by the pipeline, the complex voltage data from the radio telescope are saved
to a disk for offline processing (Section~\ref{sec:buff}).
When the false positive rate is too high, the pipeline latency can be large enough such that
it will not save the complex voltage data in time to do any further analysis. Furthermore, too many hits will lead to
too many candidates and too much voltage data filling up the storage capacity on the machine. Artificial intelligence is
an important tool to rapidly classify real candidates from the spurious ones with excellent precision. Very recently,
convolutional neural networks (CNNs) trained over extensive datasets from several radio telescopes have been implemented
on FRB search \mbox{pipelines~\citep{2018AJ....156..256C,agarwal2020,bhat2023}.} These algorithms are extremely effective in
identifying real FRBs among the vast amount of candidates resulting from the detection pipeline. 

\subsection{Buffering, Triggering, and Alerting}
\label{sec:buff}
The scientific potential of a detection is  maximised if the bursts are localised precisely, i.e.,
pin-pointed 
to (sub-)arcsecond precision, enabling identification of the FRB host galaxy.
This determination 
is crucial for
a number of reasons.  Firstly, it allows one to accurately estimate the redshift of the FRB, thus enabling estimation of
the exact  electron column depth between us and the source, which is paramount for estimating the baryon content of the
\mbox{Universe~\citep{macquart_2020_natur}. }Secondly, it allows further follow-up of the host galaxy and the environs of the FRB, providing insight into
the progenitor~\citep{tendulkar_2017_apjl}. 
In order to do this, FRB search pipelines are often equipped with a mode whereby the discovery of the FRB triggers the
buffering of the raw complex voltage data from the telescope. Due to the large amount of space that these data take,
only the data around the time of the FRB are saved to the disk. A trigger announcing the discovery of a FRB
\citep{2017arXiv171008155P}
can further be
broadcast over the internet to other radio telescopes (e.g.~\citep{2021Natur.596..505P}) and multi-wavelength facilities in order to facilitate prompt
follow-up.
When using an interferometer, images of the sky can be made from the buffered data to localise the FRB to
arc-second precision. The data can also be used to study the radio emission in the FRB at high time resolution, to reveal finer structure and components. 

%
\section{FRBs in the Era of SKAO/ngVLA}
The radio astronomy landscape will be further transformed in the coming years.
On the one hand,
new, more advanced radio telescopes -- the Square Kilometre Array (SKA)~\citep{skao}, the next-generation Very Large Array (ngVLA)~\citep{ngvla}, the Deep Synoptic Array (DSA-2000)~\citep{dsa2000},
and the Canadian Hydrogen Observations and Radio-transient Detector (CHORD)~\citep{chord}---will bring about an exponential increase in
the data volume that will need to be processed in order to detect and localise FRBs. Compounding this challenge, on the other hand, is the recent realisation that the  parameter
  space populated by FRBs may be larger than what we have been searching so-far.
Some of the existing FRB detection algorithms may not be able to keep up with such high
data rates of multiple TBs per second over  large ranges of search parameters.
In this section we first describe the opportunities for searching outside the existing box and next
  look at newer, faster, and more efficient algorithms.
We note that some of these algorithms have already been tested on a conceptual level and may be deployed
on these next-generation telescopes, as their FRB search software stack is not yet set in stone.
Here, we look a number of possible alternatives for future FRB searches with next-generation facilities. We note that this by no means is an exhaustive list and describe the ones that we find the most promising. 

\subsection{Expanding the FRB Parameter Space}
\begin{figure}[H]
\begin{adjustwidth}{-\extralength}{0cm}
\centering
\includegraphics[width=15cm]{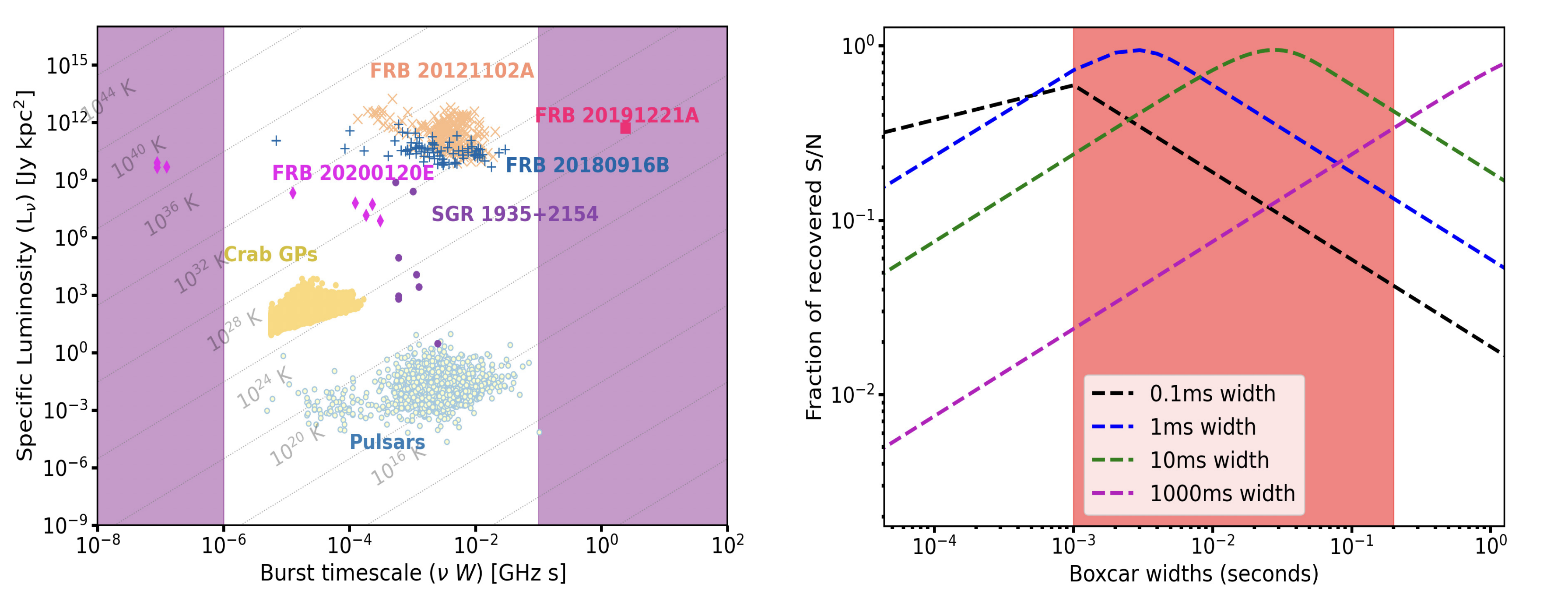}
\end{adjustwidth}
\caption{(\textbf{Left panel}) the radio transient phase space showing the spectral duration (product of the duration and frequency of emission) of coherent emitters compared to the specific luminosity (luminosity integrated over a finite radio band) of the coherent emission with two shaded regions that are mostly unexplored. The parallel lines denote the minimum specific luminosity for a given brightness temperature as a function of spectral duration. (\textbf{Right panel}) the fraction of recovered S/N as a function of boxcar width. The shaded region shows the typical range of boxcar widths used in conventional FRB searches.}\label{fig:tps}
\end{figure}

Searches for FRBs and radio pulses from neutron stars are based on the assumption that the pulses are narrow
(between 0.1--100 ms). 
Due to computational constraints, box-car widths used for match filtering never go beyond a few hundred
ms. This notion has recently required a fresh new look, thanks to the discovery of the FRB20191221A~\citep{2022Natur.607..256C} and its quasi-periodic burst components that span a total duration of as much as 3
seconds. This begs a question: if there is a significant population of wide FRBs, how many do current pipelines miss due
to width--search limitations? The right panel of Figure~\ref{fig:tps}  shows that one would miss over  40$\%$ of the burst
S/N due to overly narrow boxcar widths in matched filtering~\citep{beniamini2023}. 
On the other side of the spectrum, ref.~\citep{snelders2023} recently reported the discovery of isolated micro-second duration
bursts from FRB20121102A.
Such 
narrow FRBs 
would be largely averaged out in the data.
Together, these recent discoveries have  brought to light a selection bias in all current time-domain searches for radio
transients.
Searches on the short timescales have generally been hampered by the  computational limitations;
on the long timescales, they have been hampered by the
effects of RFI and system stability. 
Even so, it is  important to  expand the FRB search parameter space to cover the narrowest and the widest
timescales as shown in the left panel Figure~\ref{fig:tps}. Future searches for FRBs with next-generation telescopes should focus on overcoming these limitations and thus eliminating these biases.

\vspace{-6pt}

\subsection{Radon and Hough Transform}
Finding far-away and dim FRBs requires an algorithm that
maximises the signal-to-noise ratio of the pulse
by integrating the incoming dynamic spectrum along the possible quadratic dispersion curves.
Such operations happen over sets of curves form a family of
mathematical transformations that include, for example, the Radon
transform~\citep{radon1917} and Hough transform~\citep{hough1962}.
Recently, experimental attempts have been made to test the feasibility of such algorithms~\citep{schmid2018, schmid2020}. A key step here is a transformation of the dynamic spectrum from a $t-\nu$ space to $t-\frac{1}{\nu^{2}}$ space. This transformation of the parameter space turns the dispersion framework into a linear problem where the
intensity $I$ can be characterised as $I(d,t)$, where $d$ is some function of the DM, and $t$ is the time measure with a certain
reference. Then, any search with a number of DM trials can be defined as a family of linear functions in the $d-t$
parameter space, at which point finding a FRB becomes a peak detection problem using these transforms. The advantage here is that finding peaks is more robust and insensitive to any RFI that does not satisfy the $\nu^{-2}$ dependence. Below, we briefly describe the Radon transform.

\begin{figure}[H]
\begin{adjustwidth}{-\extralength}{0cm}
\centering
\includegraphics[scale=0.38]{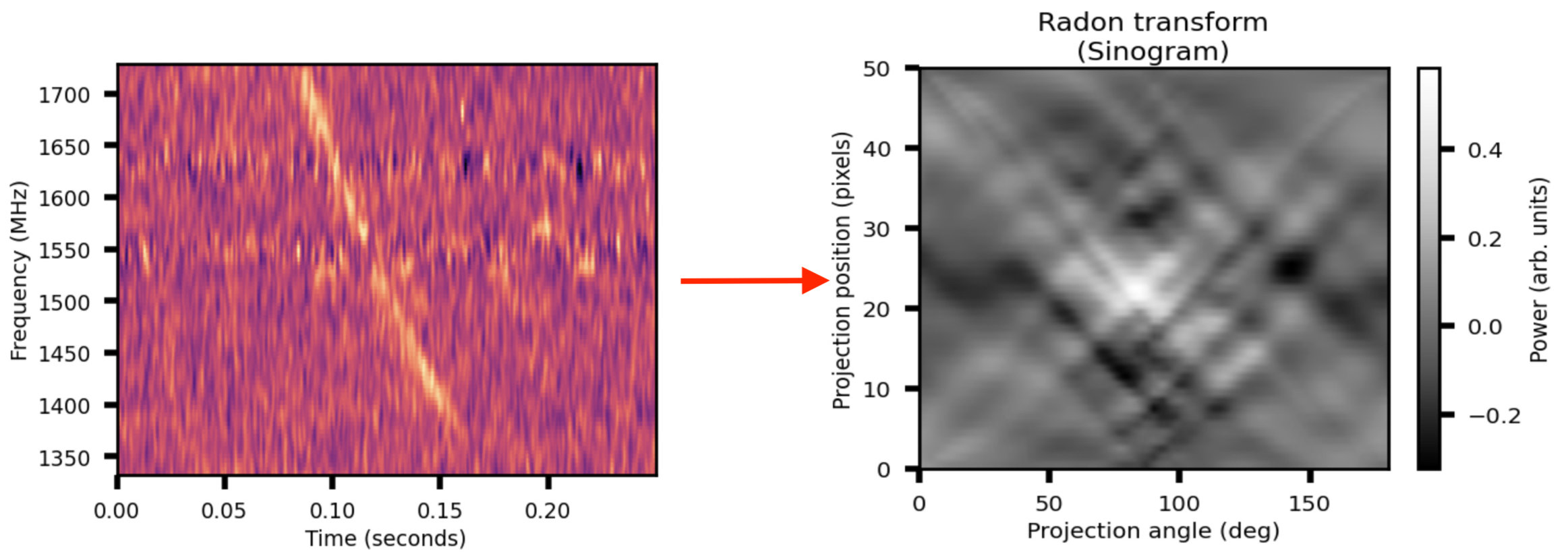}
\end{adjustwidth}
\caption{(\textbf{Left panel}) dynamic spectrum of a single pulse. (\textbf{Right panel}) corresponding Sinogram after Radon transform of the pulse showing a clear signal due to the pulse. \label{fig:radon}}
\end{figure}

Mathematically, the Radon transform is an integral transform that takes any function and projects it in a 2D space of
lines in a plane. This output  consists of the line integrals along the paths in a certain direction.  Since it is a
projection transform, it has found a wide variety of medical applications. Intriguingly, one can also use a Radon
transform to convert the received dynamic spectrum from a radio telescope to a projection space where a dispersed FRB
can be approximated as a line with some arbitrary slope. We do note that this conversion would require resampling and interpolation of the data which would also consume computational resources. These steps can be made faster with the use of fast interpolation algorithms like the nearest neighbour algorithm. A Radon transform is highly sensitive to any linear features in
the parameters space. This means that one can transform the dynamic spectrum (function of time $t$ and frequency $\nu$)
to another parameter space (a function of time $t$ and $\nu^{2}$) where any dispersed signal would be a linear
function. The position and angle of that function then produce a singular peak, as shown in Figure~\ref{fig:radon}.
These Radon transforms can then be used to search for astrophysical signals or could be served
to a neural network to classify FRBs.

\subsection{Quantum Computing Searches}
Recent advances in quantum computing justify a brief look at their potential application in FRB searches,
even if these will only be realised in the far future.
FRB searches with radio telescopes have two traits that make application of quantum algorithms highly interesting. The first is that, after the receivers record the initial voltages,
the subsequent beam forming, dedispersion, and box car searching are inherently highly parallel.
This lends itself well to encoding the respective
directions, dispersion measures, and pulse widths as
entangled states.
Then, the very large number of currently required identical classical operations---say, 1000 tied-array beams $\times$ 10,000 trial DMs $\times$ 10 boxcar widths---can be replaced by single quantum operations.
The second trait is that this large amount of data can, in the end,
be meaningfully reduced to a binary classification operation:
there either is or is not a FRB in the data.
Such collapse operations are highly suited for quantum computers.
Once detected, the FRB can be characterised fully using the relevant buffer~data. 
 
Some initial exploration of quantum computing for parallel radio-telescope operations has recently been performed
\citep{2023arXiv231012084B}. Although that work focuses on 
 image-domain data, we see analogies to the time-domain data processing required for FRB searching.
While the classification part itself is not computationally expensive,
first quantum computing steps have recently been reported for a pulsar dataset
\citep{2021arXiv211202655K} where the candidates parameters were still produced classically. 

Real-life application will require technology to convert the voltages into qubits,
algorithms that can separate the FRBs from RFI,
and quantum computers with a much larger qubit number and lower noise than will be available even in the near future.
Based on manufacturer road maps these may arguably only materialise in the 2040s.

\section{Concluding Remarks}
Searches for FRBs have been an important aspect of time-domain radio astronomy in the current decade---and they will continue to be in the upcoming one.
Finding and studying ever more FRBs 
will only accelerate the speed of progress in the field and take us closer to solving key open questions in
cosmology. Faster and more efficient search algorithms will allow us to keep up with the ever-increasing data volumes
from next-generation radio telescopes.
Expanding the FRB search parameter space 
will reveal new insights into the mysterious physics that creates them.
Development of new search tools along with state-of-the art detection systems powered by artificial intelligence will be the backbone of FRB science in the coming years.  The future looks bright for FRB science.


%




\vspace{6pt} 



\authorcontributions{K.M.R. and J.v.L.; methodology, K.M.R. and J.v.L.; formal analysis, K.M.R.; resources, K.M.R.; data curation, K.M.R.; writing—original draft preparation, K.M.R. and J.v.L.; writing—review and editing. All authors have read and agreed to the published version of the~manuscript.}

\funding{This research received support from the Vici research project ARGO (project number 639.043.815), and from NWA-ORC project CORTEX (NWA.1160.18.316). Both are financed by the Dutch Research Council (NWO) with PI J.v.L.} 

\dataavailability{The original contributions presented in the study are included in the article/supplementary material, further inquiries can be directed to the corresponding author. The raw data presented in this article will be made available by the authors on reasonable request.} 

\acknowledgments{K.M.R and J.v.L. would like to thank the reviewers whose feedback greatly improved the quality of the manuscript. K.M.R. acknowledges Barak Zackay for kindly providing the written consent to reproduce a Figure from their previous work. The authors would like to thank Duncan Lorimer, Leon Oostrum and Manisha Caleb for useful feedback that significantly improved the manuscript.  
}

\conflictsofinterest{The authors declare no conflicts of interest. The funders had no role in the design of the study; in the collection, analyses, or interpretation of data; in the writing of the manuscript; or in the decision to publish the~results.} 





\appendixtitles{yes} 

\appendixstart
\appendix
\section[\appendixname~\thesection]{Resources}
\label{app1}
The following section 
 provides links and resources for FRB search software that is publicly available: 

\begin{table}[H]
 \newcolumntype{C}{>{\centering\arraybackslash}X}
    \begin{tabularx}{\textwidth}{Ll}
        \toprule
\multicolumn{2}{l}{RFI Excision Software}\\
\midrule
IQRM & \url{https://gitlab.com/kmrajwade/iqrm_apollo}\\
RFIClean & \url{https://github.com/ymaan4/RFIClean}\\
RFim & \url{https://github.com/TRASAL/RFIm}\\
\midrule
\multicolumn{2}{l}{Dedispersion + Matched Filtering software}\\
\midrule
Heimdall & \url{https://sourceforge.net/projects/heimdall-astro/}\\
Astro-Accelerate & \url{https://github.com/AstroAccelerateOrg/astro-accelerate}\\
FDMT & \url{https://github.com/kiranshila/FastDMTransform.jl}\\
\midrule
\multicolumn{2}{l}{Candidate classification software}\\
\midrule
FETCH & \url{https://github.com/devanshkv/fetch}\\
FRBID &  \url{https://github.com/Zafiirah13/FRBID} \\
\midrule
\multicolumn{2}{l}{Full stack FRB search pipeline (all of the above)}\\
\midrule
AMBER  &  \url{https://github.com/TRASAL/AMBER}\\
TransientX & \url{https://github.com/ypmen/TransientX}\\
PRESTO & \url{https://www.cv.nrao.edu/~sransom/presto/}\\
\bottomrule
    \end{tabularx}
\end{table}



\begin{adjustwidth}{-\extralength}{0cm}
\printendnotes[custom] 

\reftitle{References}

\PublishersNote{}
\end{adjustwidth}
\end{document}